\documentclass[aps,pre,twocolumn,showpacs]{revtex4}%
\usepackage{amsfonts}
\usepackage{amsmath}
\usepackage{amssymb}
\usepackage{graphicx}%
\begin{document}
\title{Diffusion of Monochromatic Classical Waves}
\author{Sijmen Gerritsen\footnote{Email address: s.gerritsen@tnw.tudelft.nl} and Gerrit
E. W. Bauer }
\affiliation{Kavli Institute of Nanoscience, Delft University of Technology, Lorentzweg 1,
2628CJ Delft, The Netherlands}


\begin{abstract}
We study the diffusion of monochromatic classical waves in a disordered
acoustic medium by scattering theory. In order to avoid artifacts associated
with mathematical point scatterers, we model the randomness by small but
finite insertions. We derive expressions for the configuration-averaged energy
flux, energy density, and intensity for one, two and three dimensional (1D, 2D
and 3D) systems with an embedded monochromatic source using the ladder
approximation to the Bethe-Salpeter equation. We study the transition from
ballistic to diffusive wave propagation and obtain results for the
frequency-dependence of the medium properties such as mean free path and
diffusion coefficient as a function of the scattering parameters. We discover
characteristic differences of the diffusion in 2D as compared to the
conventional 3D case, such as an explicit dependence of the energy flux on the
mean free path and quite different expressions for the effective transport velocity.

\end{abstract}
\received{\today{}}

\maketitle

\section{Introduction}

The ongoing interest in the field of classical waves in complex media is
caused by the importance of detection and imaging techniques that are based on
wave propagation and scattering. This ranges from electromagnetic waves in
optical and near infrared tomography \cite{ar1} and microwave radars
\cite{fu1} to acoustic waves in ultrasonics \cite{fr1} and geophysics
\cite{sn1}. Complexity is often associated with inhomogeneities that cause
scattering which complicates most imaging processes considerably. However,
when used cleverly, the scattered field can also be used to improve imaging
\cite{ld1}. Although length scales (with respect to the wavelength) and the
degree of the disorder may vary considerably from field to field, methods and
results have been shown to be interchangeable without much difficulty
\cite{ps1}. Recent topics of interest include localization of classical waves
\cite{wl1,hs1}, the transition from ballistic to diffusive wave propagation
\cite{zj1}, acoustic time-reversal imaging \cite{mf1}, etc. Direct simulation
by the exact solution of a well-known Helmholtz wave equation for a given
realization of the medium is often the method of choice for given
applications. The drawbacks of the brute force computational approach are the
limited system size and statistics that can be achieved with given computer
resources as well as the difficulty to distill general principles out from the
plethora of output data. The need for simple models with transparent results
therefore remains.

An analytic theory of wave propagation in disordered media necessarily relies
on simple model scatterers, for which point scatterers, i.e. (regularized)
$\delta$-functions in real space, are often chosen \cite{vc1,rn1}.
Unfortunately, the scattering response of a single point scatterer can become
non-causal, a pathological behavior that can not be solved by a simple
momentum cutoff \cite{bf1}. Especially for the study of the frequency
dependence over a wider range it is therefore necessary to use more realistic
model scatterers.

In this paper we wish to study a simple but not unrealistic experiment for the
determination of the scattering properties of scalar waves in a disordered
bulk material. A signal is emitted by a source and detected by a receiver,
both embedded in the medium at sufficiently large distances from the
boundaries. Ultimately, we are interested in the detector signal caused by a
pulsed (broadband) signal emitted by the source. After a first arrival we then
expect the so-called coda that arrives at later times due to multiple
scattering at the random scatterers \cite{sf1}. However, combining both the
effects of multiple scattering and the full frequency dependence of the
scattering processes renders an analytical treatment difficult without
additional approximations, such as a complete neglect of the frequency
dependence of the scattering amplitudes \cite{tf1}. In order to understand how
to justify certain approximations and eventually find better ones, we have
carried out a study of the frequency dependence of the scattering properties
of random media. We concentrate on the steady state in the presence of
strictly monochromatic sources, which distinguishes the present work from
related studies of the propagation of narrow band pulses \cite{lt1}. As main
results we obtain the frequency dependence of macroscopic effective medium
properties like the mean free path and the diffusion constant that depend on
the microscopic parameters of the random scatterers.

When the ratio between source-receiver distance and mean free path is small,
wave propagation is predominantly ballistic. When this ratio is large, energy
and intensity propagation is governed by the diffusion equation \cite{rn1,tf1}%
. Both these regimes are well understood. However, many imaging applications
operate on length scales where the mean free path and the source-receiver
distance are comparable. This is especially the case in geophysics where mean
free paths range from a few hundred meters up to tens of kilometers
\cite{ht1}. The behavior at this crossover regime between ballistic and
diffuse wave (intensity) propagation is of considerable interest \cite{zj1}
and also subject of the present study.

Here we present an analytical formalism on monochromatic wave intensity and
energy propagation in one dimensional, two dimensional and three dimensional
(1D, 2D and 3D) homogeneously disordered media using realistic model
scatterers. We determine the relative contributions of diffusively and
coherently propagated waves as the source-receiver distance increases. We did
no find many theoretical studies of wave propagation in two dimensional random
media in the literature \cite{ps1,td1}, although several experiments on quasi
2D systems have been carried out \cite{tf1,st1}. Another possible test for our
2D theory is comparison with numerical studies, which for very large systems
are much cheaper than in the 3D case.

The remainder of this paper is organized as follows. In sections II-IV we
start with defining our model system and the basic equations, addressing the
scattering matrices of single scatterers and discussing the average amplitude
propagators in the frequency domain. The intensity, energy flux and energy
density are discussed in section V. Results on the frequency dependence of the
diffusion constant and its dependence on the model parameters are discussed in
section VI. In general, the results for 1D systems are easily obtained,
whereas our results for 3D systems agree with findings previously reported by
others. The mathematics in the 2D case is not trivial, however, and the
derivations are summarized in the appendix. We end with the conclusions.

\section{Definitions and basic equations}

\subsection{Microscopic equations}

We describe the propagation of (scalar) acoustic waves in a microscopic model
system. Specifically, we consider a 1D, 2D or 3D acoustic medium with wave
velocity $c_{0}$ and a mass density $\rho_{0}$. The medium contains $n$
randomly distributed scatterers per unit length, area or volume and we treat
the dilute limit in which the average distance between scatterers is much
larger than their radius $a$. The internal wave velocity of a scatterer is
$c_{int}$ and, for simplicity, the difference in mass density with the
surrounding medium is disregarded. The waves are emitted by a monochromatic
point source oscillating at frequency $\omega$ positioned at the origin. The
wave amplitude $\psi_{\omega}$, related to the hydrostatic pressure by
$p_{\omega}=\partial_{t}\psi_{\omega}$ and to local particle velocity by
$v_{\omega}=-\rho_{0}^{-1}\nabla\psi_{\omega}$, then obeys the wave equation:%
\begin{equation}
\left(  \mathbf{\nabla}^{2}+c^{-2}\left(  \mathbf{r}\right)  \partial_{t}%
^{2}\right)  \psi_{\omega}\left(  \mathbf{r};t\right)  =-Q\rho_{0}%
\delta\left(  \mathbf{r}\right)  \cos\left(  \omega t\right)  \text{.}
\label{WavEq}%
\end{equation}
The source term chosen here corresponds to a volume injection term. The source
emits plane waves for 1D, cylindrical waves for 2D and spherical waves for 3D
media. In all cases $Q$ is in units of length per unit time. The wave velocity
profile of the entire medium $c\left(  \mathbf{r}\right)  $ contains the
information of the positions of the scatterers (in 1D $\mathbf{r}=x$).

The Green function of the Helmholtz equation (\ref{WavEq}) in the real space
and frequency domain reads%

\begin{equation}
\left[  \mathbf{\nabla}^{2}+\kappa_{0}^{2}-V\left(  \mathbf{r};\omega\right)
\right]  G\left(  \mathbf{r},\mathbf{r}^{\prime};\omega\right)  =\delta\left(
\mathbf{r-r}^{\prime}\right)  \text{,}%
\end{equation}
where $\kappa_{0}=\omega/c_{0}$, the length of the wave vector in the
homogeneous medium. $V\left(  \mathbf{r};\omega\right)  $\ is the scattering
or impurity potential, a sum over all individual scattering potentials:%
\begin{equation}
V\left(  \mathbf{r};\omega\right)  =\kappa_{0}^{2}\left(  1-\gamma
^{-2}\right)
%TCIMACRO{\dsum \limits_{i=1}^{N}}%
%BeginExpansion
{\displaystyle\sum\limits_{i=1}^{N}}
%EndExpansion
\Theta\left(  a-\left\vert \mathbf{r}-\mathbf{r}_{i}\right\vert \right)
\text{.} \label{ScatPot}%
\end{equation}
$\Theta$ is the Heaviside step function, with $\Theta\left(  x\right)  =0$
when $x<0$ and $1$ otherwise. The velocity contrast is defined as
$\gamma=c_{int}/c_{0}$ so that the single scatterer potential is
\textquotedblleft attractive\textquotedblright\ when $\gamma<1$ and
\textquotedblleft repulsive\textquotedblright\ when $\gamma>1$. Eq.
(\ref{ScatPot}) describes a spherical potential, but the precise shape is not
relevant when the scatterers are sufficiently small compared to the wave length.

The amplitude of the wave field is related to the Green function as%
\begin{equation}
\psi_{\omega}\left(  \mathbf{r};t\right)  =-Q\rho_{0}\operatorname{Re}\left\{
e^{-i\omega t}G\left(  \mathbf{r},\mathbf{r}^{\prime}=0;\omega\right)
\right\}  \text{.}%
\end{equation}
The intensity $I_{\omega}\left(  \mathbf{r};t\right)  $ is the square of this
expression. Related physical properties are the energy flux:%
\begin{equation}
\mathbf{F}_{\omega}\left(  \mathbf{r};t\right)  =-\frac{1}{\rho_{0}}%
\partial_{t}\psi_{\omega}\left(  \mathbf{r};t\right)  \mathbf{\nabla}%
\psi_{\omega}\left(  \mathbf{r};t\right)  \text{,}%
\end{equation}
and the energy density:%
\begin{equation}
W_{\omega}\left(  \mathbf{r};t\right)  =\frac{1}{2\rho_{0}}\left(  \left(
\mathbf{\nabla}\psi_{\omega}\left(  \mathbf{r};t\right)  \right)  ^{2}%
+c^{-2}\left(  \mathbf{r}\right)  \left(  \partial_{t}\psi_{\omega}\left(
\mathbf{r};t\right)  \right)  ^{2}\right)  \text{,}%
\end{equation}
recognized as the sum of the potential and kinetic energy contributions,
respectively. For a monochromatic source with frequency $\omega$ these
observables contain a time independent contribution and a second term
oscillating with frequency $2\omega$. We concentrate on the constant part by
time-averaging over one period. Expressed in terms of the Green function this
yields%
\begin{equation}
I_{\omega}\left(  \mathbf{r}\right)  =\frac{Q^{2}\rho_{0}^{2}}{2}\left\vert
G\left(  \mathbf{r},\mathbf{r}^{\prime}=0;\omega\right)  \right\vert
^{2}\text{,} \label{IntGr}%
\end{equation}%
\begin{equation}
\mathbf{F}_{\omega}\left(  \mathbf{r}\right)  =-\frac{Q^{2}\rho_{0}\omega}%
{2}\operatorname{Im}\left\{  G\left(  \mathbf{r},\mathbf{r}^{\prime}%
=0;\omega\right)  \mathbf{\nabla}G^{\ast}\left(  \mathbf{r},\mathbf{r}%
^{\prime}=0;\omega\right)  \right\}  \text{,}%
\end{equation}%
\begin{align}
W_{\omega}\left(  \mathbf{r}\right)   &  =\frac{Q^{2}\rho_{0}}{4}\left(
\left\vert \mathbf{\nabla}G\left(  \mathbf{r},\mathbf{r}^{\prime}%
=0;\omega\right)  \right\vert ^{2}\right. \nonumber\\
&  \text{ \ \ \ \ \ \ \ \ \ \ \ \ }+\frac{\omega^{2}}{c^{2}\left(
\mathbf{r}\right)  }\left.  \left\vert G\left(  \mathbf{r},\mathbf{r}^{\prime
}=0;\omega\right)  \right\vert ^{2}\right)  \text{.}%
\end{align}

\subsection{Macroscopic equations}

The properties of the wave field depend via the Green function on the exact
configuration of scatterers. However, in large systems, different realizations
of the ensemble give similar responses (ergodicity). The similarities in the
response can be studied by calculating the configurational average. This
average is the connection between the microscopic description and the
macroscopic (effective) medium properties.

The macroscopic (diffusively scattered) intensity of pulsed sources is usually
described by the diffusion equation:%
\begin{equation}
\partial_{t}\left\langle I\left(  \mathbf{r};t\right)  \right\rangle
=D\nabla^{2}\left\langle I\left(  \mathbf{r};t\right)  \right\rangle \text{,}
\label{DiffEq}%
\end{equation}
where the brackets denote the configuration average and $D$ is the diffusion
constant. In spite of neglecting the frequency dependence of the scattering
processes, this approximation is known to work well in certain cases (under
the condition that the source receiver distance is much larger than the mean
free path) \cite{rn1,tf1}.

In order to obtain the steady-state diffuse intensity of a monochromatic wave
field, the diffusion equation (\ref{DiffEq}) is not sufficient. The
\textit{energy density (}and not the intensity) of the wave field is the
conserved property. Eq. (\ref{DiffEq}) is therefore only valid if the
intensity is strictly proportional to the energy density. In general, the
averaged energy transport is governed by Fick's first law:%
\begin{equation}
\left\langle \mathbf{F}_{\omega}\left(  \mathbf{r}\right)  \right\rangle
=-D\left(  \omega\right)  \nabla\left\langle W_{\omega}\left(  \mathbf{r}%
\right)  \right\rangle \text{,} \label{Fick1}%
\end{equation}
accounting for the frequency dependence of the diffusion constant. In the
steady-state problem and outside the monochromatic source the proper Laplace
equation is%
\begin{equation}
\nabla^{2}\left\langle W_{\omega}\left(  \mathbf{r}\right)  \right\rangle
=0\text{.}%
\end{equation}

\section{Scattering matrices}

Here we discuss the properties of a single model scatterer in the system
($N=1$ in equation (\ref{ScatPot})). The response of a system containing a
monochromatic source (in the origin), a receiver (at $\mathbf{r}$) and a
single \textquotedblleft s-wave\textquotedblright\ scatterer (at
$\mathbf{r}_{i}$) can be expressed in terms of Green functions of the
homogeneous system ($V=0$) \cite{am1}:%
\begin{align}
G\left(  \mathbf{r},\mathbf{r}^{\prime}=0;\omega\right)   &  =G_{0}\left(
r;\omega\right) \nonumber\\
&  +G_{0}\left(  \left\vert \mathbf{r}-\mathbf{r}_{i}\right\vert
;\omega\right)  t_{0}\left(  \omega\right)  G_{0}\left(  r;\omega\right)
\text{.}%
\end{align}
This expression is valid in the far field limit\ ($r,r_{i}\gg\lambda$) and
when scattering is isotropic ($\lambda\gg a$), where $\lambda$ is the wavelength.

The transition (t-) matrix elements for s-wave scattering are related to the
scattering matrix elements by%
\begin{equation}
t_{0}\left(  \omega\right)  =\left\{
\begin{array}
[c]{c}%
2i\kappa_{0}R\left(  \omega\right)  \text{ \ \ \ \ \ \ \ \ \ \ \ \ \ \ \ (1D)}%
\\
2i\left(  S_{0}\left(  \omega\right)  -1\right)  \text{
\ \ \ \ \ \ \ \ \ (2D)}\\
2\pi i\kappa_{0}^{-1}\left(  S_{0}\left(  \omega\right)  -1\right)  \text{
\ \ (3D)}%
\end{array}
\right.  \text{.}%
\end{equation}
In 1D, the s-wave scattering condition corresponds to equivalence of $t_{0}$
for either reflection or transmission. $R\left(  \omega\right)  $ is the
reflection coefficient at a step discontinuity, it can be obtained by imposing
flux conservation across the scatterer boundary. This gives \cite{em1}%
\begin{equation}
R\left(  \omega\right)  =e^{-i\kappa_{0}2a}\frac{R_{0}\left(  1-e^{i\kappa
_{0}4a/\gamma}\right)  }{1-R_{0}^{2}e^{i\kappa_{0}4a/\gamma}}\text{,}%
\end{equation}
where $R_{0}=\left(  \gamma-1\right)  /\left(  \gamma+1\right)  $. In the same
way we can derive an expression for the scattering matrix element of the
s-wave channel $S_{0}$ (related to the scattering phase shift $\delta_{0}$ by
$S_{0}=\exp\left(  i2\delta_{0}\right)  $). In 2D \cite{es1}%
\begin{align}
&  S_{0}\left(  \omega\right) \nonumber\\
&  =-\frac{\gamma J_{0}\left(  \kappa_{0}a/\gamma\right)  H_{1}^{\left(
2\right)  }\left(  \kappa_{0}a\right)  -J_{1}\left(  \kappa_{0}a/\gamma
\right)  H_{0}^{\left(  2\right)  }\left(  \kappa_{0}a\right)  }{\gamma
J_{0}\left(  \kappa_{0}a/\gamma\right)  H_{1}^{\left(  1\right)  }\left(
\kappa_{0}a\right)  -J_{1}\left(  \kappa_{0}a/\gamma\right)  H_{0}^{\left(
1\right)  }\left(  \kappa_{0}a\right)  }\text{.}%
\end{align}
In 3D the Bessel ($J_{i}$) and Hankel ($H_{i}^{\left(  j\right)  }$) functions
are replaced by the spherical Bessel ($j_{i}$) and Hankel ($h_{i}^{\left(
j\right)  }$) functions. The scattering matrix element then simplifies to
\cite{em1}%
\begin{equation}
S_{0}\left(  \omega\right)  =e^{-i2\kappa_{0}a}\frac{\cot\left(  \kappa
_{0}a/\gamma\right)  +i\gamma}{\cot\left(  \kappa_{0}a/\gamma\right)
-i\gamma}\text{.}%
\end{equation}

\section{The configuration-averaged propagator}

Now we switch to the case of multiple scattering at the proposed model
scatterers. The wave propagator in a disordered medium after configuration
averaging is dressed with a self-energy $\Sigma$. In reciprocal space it reads
\cite{rn1}%
\begin{equation}
\left\langle G\left(  \mathbf{k,k}^{\prime};\omega\right)  \right\rangle
=\frac{1}{\kappa_{0}^{2}-\mathbf{k}^{2}-\Sigma\left(  \mathbf{k}%
;\omega\right)  }\left(  2\pi\right)  ^{d}\delta^{\left(  d\right)  }\left(
\mathbf{k}-\mathbf{k}^{\prime}\right)  \text{,} \label{AvAmpk}%
\end{equation}
where $d$ is the dimension and $\delta^{\left(  d\right)  }$ the Dirac delta
function. When $n$, the density of scatterers, is low, interference between
multiply scattered waves by different sites may be disregarded. In this
\textquotedblleft single site approximation\textquotedblright\ the self-energy
does not depend on $\mathbf{k}$ and it is simply given by \cite{rn1}%
\begin{equation}
\Sigma\left(  \omega\right)  =nt_{0}\left(  \omega\right)  \text{.}
\label{SelfEnt0}%
\end{equation}

This approximation does not restrict the scattering strength since $t_{0}$\ is
the full scattering matrix of the single scatterer. Interference effects from
multiple scattering at different scatterers cause localization that is known
to be important in 1D (where the localization length is of the order of the
mean free path) and in 2D media (where the localization length is a
transcendental function of the mean free path). In 3D, localization can be
disregarded except for very strong scattering media \cite{wl1}. Here we
restrict ourselves to purely non-localized transport phenomena, remembering
that we can always find a region where this type of transport is dominant.

Fourier transforming (\ref{AvAmpk}) with self-energy given by (\ref{SelfEnt0})
to real space gives the averaged Green function that depends only on the
source-receiver distance ($G\left(  r;\omega\right)  =\left\langle G\left(
\mathbf{r},\mathbf{r}^{\prime}=0;\omega\right)  \right\rangle $). In 1D the
amplitude propagators are exponentially damped plane waves:%
\begin{equation}
G\left(  \left\vert x\right\vert ;\omega\right)  =\frac{1}{2i\kappa_{e}\left(
\omega\right)  }e^{i\kappa_{e}\left(  \omega\right)  \left\vert x\right\vert
}\text{,} \label{Grw1D}%
\end{equation}
in 2D they are cylindrical:%
\begin{equation}
G\left(  r;\omega\right)  =\left\{
\begin{array}
[c]{c}%
-\frac{i}{4}H_{0}^{(1)}\left(  \kappa_{e}\left(  \omega\right)  r\right)
\text{ \ \ if \ }\omega>0\\
\frac{i}{4}H_{0}^{(2)}\left(  -\kappa_{e}\left(  \omega\right)  r\right)
\text{ \ \ if \ }\omega<0
\end{array}
\right.  \text{'} \label{Grw2D}%
\end{equation}
and in 3D spherical:%
\begin{equation}
G\left(  r;\omega\right)  =\frac{-1}{4\pi r}e^{i\kappa_{e}\left(
\omega\right)  r} \label{Grw3D}%
\end{equation}
\cite{ps1}. In Eq. (\ref{Grw1D}-\ref{Grw3D}) $\kappa_{e}$ is the
\textquotedblleft renormalized\textquotedblright\ effective wave vector:%
\begin{align}
\kappa_{e}\left(  \omega\right)   &  =\sqrt{\kappa_{0}^{2}-nt_{0}\left(
\omega\right)  }\nonumber\\
&  \equiv sgn\left(  \omega\right)  \kappa_{r}\left(  \omega\right)
+i\frac{1}{2\ell_{f}\left(  \omega\right)  }\text{.}%
\end{align}
$\kappa_{r}\left(  \omega\right)  =\left\vert \operatorname{Re}\kappa
_{e}\left(  \omega\right)  \right\vert $ and $\ell_{f}^{-1}\left(
\omega\right)  =2\left\vert \operatorname{Im}\kappa_{e}\left(  \omega\right)
\right\vert $, the mean free path. We retrieve the Green functions for the
homogeneous systems ($G_{0}$) by letting $n$ or $t_{0}$ go to zero. Properties
of the averaged response to a pulsed signal can be studied by calculating the
Fourier transform to the time domain, as was done in Refs. \cite{bf1} and
\cite{fb1}.

\section{The\ configuration-averaged\ intensity end energy}

We derive here the configuration averaged intensity, energy flux and energy
density in the frequency domain.

\subsection{The Bethe-Salpeter equation}

Ensemble averaging the intensity (\ref{IntGr}) gives us%
\begin{equation}
\left\langle I_{\omega}\left(  \mathbf{r}\right)  \right\rangle =\frac
{Q^{2}\rho_{0}^{2}}{2}\Pi\left(  r;\omega\right)  \text{,}%
\end{equation}
where $\Pi\left(  r;\omega\right)  =\left\langle \left\vert G\left(
\mathbf{r},\mathbf{r}^{\prime}=0;\omega\right)  \right\vert ^{2}\right\rangle
$ is the average of the squared Green function propagator. It is given by%
\begin{align}
\Pi\left(  r;\omega\right)   &  =\Pi_{0}\left(  r;\omega\right)  +%
%TCIMACRO{\dint }%
%BeginExpansion
{\displaystyle\int}
%EndExpansion
d^{d}\mathbf{r}_{1}d^{d}\mathbf{r}d^{d}\mathbf{r}_{3}d^{d}\mathbf{r}%
_{4}\left\langle G\left(  \mathbf{r},\mathbf{r}_{1};\omega\right)
\right\rangle \nonumber\\
&  \times\left\langle G^{\ast}\left(  \mathbf{r},\mathbf{r}_{2};\omega\right)
\right\rangle \Gamma\left(  \mathbf{r}_{1},\mathbf{r}_{2},\mathbf{r}%
_{3},\mathbf{r}_{4};\omega\right) \nonumber\\
&  \times\left\langle G\left(  \mathbf{r}_{3},\mathbf{r}^{\prime}%
=0;\omega\right)  G^{\ast}\left(  \mathbf{r}_{4},\mathbf{r}^{\prime}%
=0;\omega\right)  \right\rangle \text{.}%
\end{align}
This is the Bethe-Salpeter equation in position\textbf{ }space, where $\Pi
_{0}$ is the coherent intensity ($\Pi_{0}=\left\vert \left\langle
G\right\rangle \right\vert ^{2}$) and $\Gamma\ $is the irreducible vertex
function. The lowest order approximation that still accounts for multiple
scattering is%
\begin{align}
\Gamma\left(  \mathbf{r}_{1},\mathbf{r}_{2},\mathbf{r}_{3},\mathbf{r}%
_{4};\omega\right)   &  =n\Gamma\left(  \omega\right)  \delta^{\left(
d\right)  }\left(  \mathbf{r}_{1}-\mathbf{r}_{3}\right) \nonumber\\
&  \times\delta^{\left(  d\right)  }\left(  \mathbf{r}_{1}-\mathbf{r}%
_{2}\right)  \delta^{\left(  d\right)  }\left(  \mathbf{r}_{3}-\mathbf{r}%
_{4}\right)  \text{.} \label{IrrVertAppr}%
\end{align}
This reduces the Bethe-Salpeter equation to%
\begin{align}
\Pi\left(  r;\omega\right)   &  =\Pi_{0}\left(  r;\omega\right) \nonumber\\
&  +n\Gamma\left(  \omega\right)
%TCIMACRO{\dint }%
%BeginExpansion
{\displaystyle\int}
%EndExpansion
d^{d}\mathbf{r}_{1}\Pi_{0}\left(  \left\vert \mathbf{r}-\mathbf{r}%
_{1}\right\vert ;\omega\right)  \Pi\left(  r_{1};\omega\right)  \text{.}%
\end{align}
In reciprocal space this integral equation becomes a geometric series that can
be summed as%
\begin{equation}
\Pi\left(  k;\omega\right)  =\frac{\Pi_{0}\left(  k;\omega\right)  }%
{1-n\Gamma\left(  \omega\right)  \Pi_{0}\left(  k;\omega\right)  }\text{.}
\label{Pik}%
\end{equation}

In order to be able to calculate the Fourier transform of $\Pi\left(
k;\omega\right)  $, an expression for $\Pi_{0}\left(  k;\omega\right)  $ is
needed. It is calculated as the Fourier transform of the coherent intensity
and this results in 1D in%

\begin{equation}
\Pi_{0}\left(  k;\omega\right)  =\frac{2\ell_{f}^{3}}{\left(  \left(
2\kappa_{r}\ell_{f}\right)  ^{2}+1\right)  \left(  \left(  k\ell_{f}\right)
^{2}-1\right)  }\text{,} \label{Pi0k1D}%
\end{equation}
in 2D in%
\begin{equation}
\Pi_{0}\left(  k;\omega\right)  =\frac{\ell_{f}^{2}}{\pi}\frac{\arcsin\left(
\frac{\sqrt{\left(  2\kappa_{r}\ell_{f}\right)  ^{2}-\left(  k\ell_{f}\right)
^{2}}}{\sqrt{1+\left(  2\kappa_{r}\ell_{f}\right)  ^{2}}}\right)  }%
{\sqrt{1+\left(  k\ell_{f}\right)  ^{2}}\sqrt{\left(  2\kappa_{r}\ell
_{f}\right)  ^{2}-\left(  k\ell_{f}\right)  ^{2}}}\text{,} \label{Pi0k2D}%
\end{equation}
and in 3D in \cite{rn1}%
\begin{equation}
\Pi_{0}\left(  k;\omega\right)  =\frac{\ell_{f}}{4\pi}\frac{\arctan\left(
k\ell_{f}\right)  }{k\ell_{f}}\text{.} \label{Pi0k3D}%
\end{equation}
The calculation of the vertex function $\Gamma$ is discussed in the next subsection.

\subsection{Energy conservation and the Ward identity}

It is well known that for a given approximation for the self-energy, one is
not free to choose the vertex correction. Here we take advantage of the flux
conservation constraint to obtain $\Gamma$ without additional microscopic
calculations. The energy flux from the monochromatic source (on average)
points outwards. In the steady state case the following condition must hold
for the averaged flux in direction $\mathbf{n}$:%
\begin{equation}
\left\langle \mathbf{n\cdot F}_{\omega}\left(  \mathbf{r}\right)
\right\rangle \propto\frac{1}{r^{d-1}}\mathbf{n}\cdot\widehat{\mathbf{r}%
}\text{,} \label{FluxCons2D3D}%
\end{equation}
where $\widehat{\mathbf{r}}$ is the unit vector in the radial direction. In 1D
this condition reads%
\begin{equation}
\left\langle F_{\omega}\left(  x\right)  \right\rangle \propto sgn\left(
x\right)  \text{.} \label{FluxCons1D}%
\end{equation}
The microscopic expression for the average energy flux is%
\begin{align}
\left\langle \mathbf{n\cdot F}_{\omega}\left(  \mathbf{r}\right)
\right\rangle  &  =-\frac{Q^{2}\rho_{0}\omega}{2}\nonumber\\
&  \times\operatorname{Im}\left\{  \left\langle G\left(  \mathbf{r}%
,\mathbf{r}^{\prime}=0;\omega\right)  \mathbf{n\cdot\nabla}G^{\ast}\left(
\mathbf{r},\mathbf{r}^{\prime}=0;\omega\right)  \right\rangle \right\}
\nonumber\\
&  =-\frac{Q^{2}\rho_{0}\omega}{2}\operatorname{Im}\left\{  \overset
{\prime\text{ \ }}{\Pi^{\mathbf{n}}}\left(  \mathbf{r};\omega\right)
\right\}  \text{,}%
\end{align}
which defines the function $\overset{\prime\text{ \ }}{\Pi^{\mathbf{n}}}$. The
vertex function is the same as for the intensity, so we can express
$\overset{\prime\text{ \ }}{\Pi^{\mathbf{n}}}$ in reciprocal space as%
\begin{equation}
\overset{\prime\text{ \ }}{\Pi^{\mathbf{n}}}\left(  \mathbf{k};\omega\right)
=\frac{\overset{\prime\text{ \ }}{\Pi_{0}^{\mathbf{n}}}\left(  \mathbf{k}%
;\omega\right)  }{1-n\Gamma\left(  \omega\right)  \Pi_{0}\left(
k;\omega\right)  }\text{.} \label{PinRec}%
\end{equation}
$\overset{\prime\text{ \ }}{\Pi_{0}^{\mathbf{n}}}\left(  k;\omega\right)  $ is
the coherent energy flux in direction $\mathbf{n}$ that is given by the
Fourier transform of%
\begin{equation}
\overset{\prime\text{ \ }}{\Pi_{0}^{\mathbf{n}}}\left(  \mathbf{r}%
;\omega\right)  =G\left(  r;\omega\right)  \mathbf{n\cdot\nabla}G^{\ast
}\left(  r;\omega\right)  \text{.}%
\end{equation}

In 2D and 3D the averaged microscopic expression for the energy flux should
match the macroscopic condition%
\begin{equation}
\operatorname{Im}\left\{  \overset{\prime\text{ \ }}{\Pi^{\mathbf{n}}}\left(
\mathbf{r};\omega\right)  \right\}  =-\frac{C}{r^{d-1}}\mathbf{n}\cdot
\widehat{\mathbf{r}}\text{,} \label{FluxCons2D3DPir}%
\end{equation}
which in reciprocal space reads%
\begin{equation}
\operatorname{Re}\left\{  \overset{\prime\text{ \ }}{\Pi^{\mathbf{n}}}\left(
\mathbf{k};\omega\right)  \right\}  =-\left(  \mathbf{n}\cdot\widehat
{\mathbf{k}}\right)  2^{d-1}\pi\frac{C}{k}\text{,} \label{FluxCons2D3DPik}%
\end{equation}
where $C$ is real and depends on frequency and the model parameters. $\Pi
_{0}\left(  k;\omega\right)  $ is an even function of $k$. We know how
$\overset{\prime\text{ \ }}{\Pi_{0}^{\mathbf{n}}}\left(  \mathbf{k}%
;\omega\right)  $ depends on $k$, as the Fourier transform in 2D reads%
\begin{align}
\overset{\prime\text{ \ }}{\Pi_{0}^{\mathbf{n}}}\left(  \mathbf{k}%
;\omega\right)   &  =-\left(  \mathbf{n}\cdot\widehat{\mathbf{k}}\right)  2\pi
i\nonumber\\
&  \times%
%TCIMACRO{\dint \limits_{0}^{\infty}}%
%BeginExpansion
{\displaystyle\int\limits_{0}^{\infty}}
%EndExpansion
drJ_{1}\left(  kr\right)  rG\left(  r;\omega\right)  \partial_{r}G^{\ast
}\left(  r;\omega\right)  \text{,}%
\end{align}
and in 3D%
\begin{align}
\overset{\prime\text{ \ }}{\Pi_{0}^{\mathbf{n}}}\left(  \mathbf{k}%
;\omega\right)   &  =-\left(  \mathbf{n}\cdot\widehat{\mathbf{k}}\right)  4\pi
i\nonumber\\
&  \times%
%TCIMACRO{\dint \limits_{0}^{\infty}}%
%BeginExpansion
{\displaystyle\int\limits_{0}^{\infty}}
%EndExpansion
drj_{1}\left(  kr\right)  r^{2}G\left(  r;\omega\right)  \partial_{r}G^{\ast
}\left(  r;\omega\right)  \text{.}%
\end{align}
The Taylor series of $\overset{\prime\text{ \ }}{\Pi_{0}^{\mathbf{n}}}\left(
\mathbf{k};\omega\right)  $ around $k=0$ only contains odd terms. So, in the
limit that $k\rightarrow0$, condition (\ref{FluxCons2D3DPik}) can only be
fulfilled by Eq. (\ref{PinRec}) when%
\begin{equation}
n\Gamma\left(  \omega\right)  =\Pi_{0}^{-1}\left(  k=0;\omega\right)  \text{.}
\label{WardId}%
\end{equation}
In 1D it is straightforward to show that condition (\ref{FluxCons1D}) can only
be fulfilled when Eq. (\ref{WardId}) is fulfilled as well. The Ward identities
are relations between self-energy and vertex corrections. We can identify Eq.
(\ref{WardId}) as the Ward identity for our problem. We now have all the
ingredients to calculate the Fourier transform of (\ref{Pik}) and
(\ref{PinRec}) to calculate the averaged intensity and energy flux respectively.

\subsection{Flux}

Using the Taylor expansions in the limit $k\rightarrow0$, we find an
expression for $C$ (from Eq.\ (\ref{FluxCons2D3DPir})) in 2D and 3D:%
\begin{equation}
C=\frac{%
%TCIMACRO{\dint \limits_{0}^{\infty}}%
%BeginExpansion
{\displaystyle\int\limits_{0}^{\infty}}
%EndExpansion
dr\operatorname{Im}\left\{  G\left(  r;\omega\right)  \partial_{r}G^{\ast
}\left(  r;\omega\right)  \right\}  r^{d}d^{-1}}{\frac{1}{2}\Pi_{0}%
^{-1}\left(  k=0;\omega\right)  \left.  \partial_{k}^{2}\Pi_{0}\left(
k;\omega\right)  \right\vert _{k=0}}\text{.} \label{CforFlux2D3D}%
\end{equation}
The average flux in 1D is obtained by directly Fourier transforming
(\ref{PinRec}):%
\begin{equation}
\left\langle F_{\omega}\left(  x\right)  \right\rangle =\frac{Q^{2}\rho
_{0}\left\vert \omega\right\vert }{8}\frac{\kappa_{r}}{\kappa_{r}%
^{2}+1/\left(  2\ell_{f}\right)  ^{2}}sgn\left(  x\right)  \text{.}
\label{AvFlux1D}%
\end{equation}
We show how to calculate $C$ in 2D case in the appendix. With the result, the
projection of the average flux becomes%
\begin{equation}
\left\langle \mathbf{n\cdot F}_{\omega}\left(  \mathbf{r}\right)
\right\rangle =\frac{Q^{2}\rho_{0}\left\vert \omega\right\vert }{8\pi^{2}%
}\frac{\arctan\left(  2\kappa_{r}\ell_{f}\right)  }{r}\mathbf{n}\cdot
\widehat{\mathbf{r}}\text{,} \label{AvFlux2D}%
\end{equation}
while in 3D it is straightforward to calculate $C$ from Eq.
(\ref{CforFlux2D3D}) and the projection of the average flux then reads%
\begin{equation}
\left\langle \mathbf{n\cdot F}_{\omega}\left(  \mathbf{r}\right)
\right\rangle =\frac{Q^{2}\rho_{0}\left\vert \omega\right\vert }{2}%
\frac{\kappa_{r}}{\left(  4\pi\right)  ^{2}r^{2}}\mathbf{n}\cdot
\widehat{\mathbf{r}}\text{.} \label{AvFlux3D}%
\end{equation}
Letting $\ell_{f}\rightarrow\infty$ ($\kappa_{r}\rightarrow\left\vert
\kappa_{0}\right\vert $) recovers the flux of a monochromatic source in an
unperturbed medium.

It is interesting to see that, in contrast to the 3D case, in 1D and 2D the
average flux depends on both the mean free path and the real part of the
effective wave vector. So the scattering mean free path limits the energy flux
in 1D and 2D, but not in 3D. In a strongly scattering 2D medium, in which the
wave energy is not (yet) localized, the dependence on the arctangent should be observable.

\subsection{Intensity}

The total average intensity is proportional to the propagator $\Pi\left(
r;\omega\right)  $, that can be obtained by calculating the Fourier transform:%
\begin{equation}
\Pi\left(  r;\omega\right)  =%
%TCIMACRO{\dint }%
%BeginExpansion
{\displaystyle\int}
%EndExpansion
\frac{d^{d}\mathbf{k}}{\left(  2\pi\right)  ^{d}}\frac{e^{i\mathbf{k}%
\cdot\mathbf{r}}}{\Pi_{0}^{-1}\left(  k;\omega\right)  -\Pi_{0}^{-1}\left(
k=0;\omega\right)  }\text{.} \label{FTInt}%
\end{equation}
$\Pi_{0}\left(  k\right)  $ is given by (\ref{Pi0k1D}) in 1D, (\ref{Pi0k2D})
in 2D and (\ref{Pi0k3D}) in the 3D case. In 1D and 2D this integral diverges
because in the steady state case with a monochromatic source, energy does not
escape fast enough to infinity due to the scatterers. This is analogous to the
fact that the Poisson equation (the diffusion equation in steady state with
source term) for a line or planar source has no well-defined solution.

The gradient of the intensity exists in all cases. In 1D it is constant and
the derivative of $\Pi\left(  x;\omega\right)  $\ is given by%
\begin{equation}
\partial_{x}\Pi\left(  x;\omega\right)  =-sgn\left(  x\right)  \frac{1}%
{4\ell_{f}}\frac{1}{\kappa_{r}^{2}+1/\left(  2\ell_{f}\right)  ^{2}}\text{.}%
\end{equation}

The gradient of $\Pi$ in 2D is expressed as an integral by%
\begin{align}
\nabla\Pi\left(  r;\omega\right)   &  =-\widehat{\mathbf{r}}%
%TCIMACRO{\dint \limits_{0}^{\infty}}%
%BeginExpansion
{\displaystyle\int\limits_{0}^{\infty}}
%EndExpansion
\frac{dk}{2\pi}\frac{k^{2}J_{1}\left(  kr\right)  }{\Pi_{0}^{-1}\left(
k;\omega\right)  -\Pi_{0}^{-1}\left(  k=0;\omega\right)  }\nonumber\\
&  =\widehat{\mathbf{r}}f\left(  r;\omega\right)  \text{,}%
\end{align}
which defines a function $f\left(  r;\omega\right)  $, that represents the
gradient in the $\widehat{\mathbf{r}}$ direction. We split it up into a
coherent ($coh$) and a \textquotedblleft totally diffusive\textquotedblright%
\ ($td$) part and a crossover correction ($cr$)%
\begin{equation}
f\left(  r;\omega\right)  =f_{coh}\left(  r;\omega\right)  +f_{td}\left(
r;\omega\right)  +f_{cr}\left(  r;\omega\right)  \text{.}%
\end{equation}
The coherent part is connected to the unscattered intensity, therefore%
\begin{align}
f_{coh}\left(  r;\omega\right)   &  =\partial_{r}\left\vert G\left(
r;\omega\right)  \right\vert ^{2}\nonumber\\
&  =-\frac{1}{8}\operatorname{Re}\left\{  \left(  \kappa_{r}+i/2\ell
_{f}\right)  H_{1}^{\left(  1\right)  }\left(  \left(  \kappa_{r}+i/2\ell
_{f}\right)  r\right)  \right.  \nonumber\\
&  \times\left.  H_{0}^{\left(  2\right)  }\left(  \left(  \kappa_{r}%
-i/2\ell_{f}\right)  r\right)  \right\}  \text{.}\label{2Dfcoh}%
\end{align}
In the appendix it is shown that%
\begin{equation}
f_{td}\left(  r;\omega\right)  =-\frac{\arctan\left(  2\kappa_{r}\ell
_{f}\right)  }{\pi^{2}2\kappa_{r}\ell_{f}}\frac{1}{r}g^{-1}\left(  2\kappa
_{r}\ell_{f}\right)  \text{,}\label{2Dftotdif}%
\end{equation}
with%
\begin{equation}
g\left(  2\kappa_{r}\ell_{f}\right)  =1-\frac{1}{\left(  2\kappa_{r}\ell
_{f}\right)  ^{2}}+\frac{1}{2\kappa_{r}\ell_{f}\arctan\left(  2\kappa_{r}%
\ell_{f}\right)  }\text{.}\label{g2D}%
\end{equation}
This part decays like $1/r$, much slower than the coherent and crossover
contributions. It is the part that describes the intensity gradient when
energy transport is completely governed by Fick's first law, that is why we
refer to this term as the \textquotedblleft totally
diffusive\textquotedblright\ part. When the total gradient is approximated by
the just the sum of the coherent and the totally diffusive contribution, the
gradient first decays exponentially until the source-receiver distance is
approximately two to three mean free paths and then the $1/r$ decay is
dominant. However, in this approximation it is neglected that close to the
source the diffusive field will look different than far away from the source.
The third term of $f$, the crossover term, describes this difference. In the
appendix it is shown that%
\begin{align}
&  f_{td}\left(  r;\omega\right)  +f_{cr}\left(  r;\omega\right)  \nonumber\\
&  =-%
%TCIMACRO{\dint \limits_{0}^{\infty}}%
%BeginExpansion
{\displaystyle\int\limits_{0}^{\infty}}
%EndExpansion
\frac{dk}{2\pi}J_{1}\left(  kr\right)  k^{2}\Pi_{sc}\left(  k;\omega\right)
\text{,}\label{2DDifIntIntegral}%
\end{align}
where%
\begin{equation}
\Pi_{sc}\left(  k;\omega\right)  =\frac{\Pi_{0}^{-1}\left(  k=0;\omega\right)
\Pi_{0}\left(  k;\omega\right)  }{\Pi_{0}^{-1}\left(  k;\omega\right)
-\Pi_{0}^{-1}\left(  k=0;\omega\right)  }\text{.}%
\end{equation}
We did not find an analytical expression for the integral
(\ref{2DDifIntIntegral}) and thus we need to evaluate it numerically. The
crossover term vanishes for $r/\ell_{f}\rightarrow0$ or $r/\ell_{f}\gg1$ and
peaks at $r/\ell_{f}\approx0.3$. Only around this value of $r/\ell_{f}$ the
gradient (in absolute value) is overestimated significantly (up to $25\%$)
when we approximate it by just the sum of coherent and \textquotedblleft
totally diffusive\textquotedblright\ terms.

\begin{figure}
[ptb]
\begin{center}
\includegraphics[
height=2.4881in,
width=3.4in
]%
{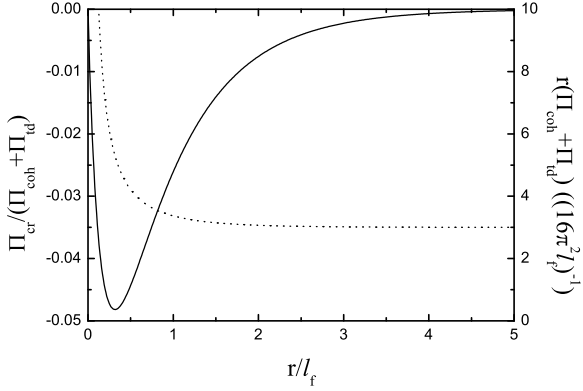}%
\caption{$r\left(  \Pi_{coh}+\Pi_{td}\right)  $ in units of $\left(  16\pi
^{2}\ell_{f}\right)  ^{-1}$ (dotted line, right axis) and $\Pi_{cr}/\left(
\Pi_{coh}+\Pi_{td}\right)  $ (solid line, left axis) as a function of
$r/\ell_{f}$ (the source-receiver distance in number of mean free paths) for
the 3D disordered medium.}%
\label{Fig3DInt}%
\end{center}
\end{figure}

In 3D the Fourier transform (\ref{FTInt}) converges and the intensity is
well-defined. We rewrite%
\begin{equation}
\Pi\left(  r;\omega\right)  =\frac{1}{16\pi^{2}r\ell_{f}}\left(  \frac
{\ell_{f}}{r}e^{-r/\ell_{f}}+3+e^{-r/\ell_{f}}h\left(  r/\ell_{f}\right)
\right)  \text{,} \label{Pirw3Dtot}%
\end{equation}
where%
\begin{align}
h\left(  r/\ell_{f}\right)   &  =%
%TCIMACRO{\dint \limits_{0}^{\infty}}%
%BeginExpansion
{\displaystyle\int\limits_{0}^{\infty}}
%EndExpansion
d\xi\left(  \frac{4\left(  \xi+1\right)  ^{2}}{\left(  2\left(  \xi+1\right)
-\ln\left(  1+2/\xi\right)  \right)  ^{2}+\pi^{2}}-1\right) \nonumber\\
&  \times e^{-\xi r/\ell_{f}}\text{.} \label{3DCroIntegral}%
\end{align}
We were not able to solve (\ref{3DCroIntegral}) analytically as well. In the
3D case, the intensity is a function of $\ell_{f}$ only (it does not depend on
$\kappa_{r}$). Eq. (\ref{Pirw3Dtot}) consists of three terms ($\Pi=\Pi
_{coh}+\Pi_{td}+\Pi\,_{cr}$). The first term is proportional to the coherent
intensity, the second term is the algebraically decaying diffuse term (the
only term that is not exponentially decaying). We plot the sum of the first
and second term multiplied by $r$ (in units of $1/\left(  16\pi^{2}\ell
_{f}\right)  $) in figure (\ref{Fig3DInt}) as a function of $r/\ell_{f}$ (on
the right axis). The third term is again the crossover correction to the total
intensity if we approximate $\Pi$ by only the first two terms. A plot of the
crossover correction divided by the sum of the first two terms is shown in
figure\ (\ref{Fig3DInt}) (left axis). The crossover term vanishes for
$r/\ell_{f}\rightarrow0$ or $r/\ell_{f}\gg1$ and peaks at $r/\ell_{f}%
\approx0.3$. It can thus be concluded that the intensity can very well be
approximated by just the sum of coherently and totally diffusively propagated
intensities, as the total intensity in 3D will never be overestimated by more
than $5\%$ using this approximation.%
%TCIMACRO{\FRAME{ftbpFU}{3.584in}{2.6227in}{0pt}{\Qcb{$r\left(  \Pi_{coh}%
%+\Pi_{td}\right)  $ in units of $\left(  16\pi^{2}\ell_{f}\right)  ^{-1}$
%(dotted line, right axis) and $\Pi_{cr}/\left(  \Pi_{coh}+\Pi_{td}\right)  $
%(solid line, left axis) as a function of $r/\ell_{f}$ (the source-receiver
%distance in number of mean free paths) for the 3D disordered medium.}%
%}{\Qlb{Fig3DInt}}{3dint.eps}{\special{ language "Scientific Word";
%type "GRAPHIC";  maintain-aspect-ratio TRUE;  display "USEDEF";
%valid_file "F";  width 3.584in;  height 2.6227in;  depth 0pt;
%original-width 4.5595in;  original-height 3.325in;  cropleft "0";
%filename '3DInt.EPS';file-properties "NPEU";}}}%
%BeginExpansion
%EndExpansion

To complete this discussion we show the final results for the gradient of the
average intensity in 1D and 2D:%
\begin{equation}
\partial_{x}\left\langle I_{\omega}\left(  x\right)  \right\rangle
=-\frac{Q^{2}\rho_{0}^{2}}{2}sgn\left(  x\right)  \frac{1}{4\ell_{f}\left\vert
\kappa_{e}\right\vert ^{2}}\text{,} \label{GrAvInt1D}%
\end{equation}%
\begin{equation}
\nabla\left\langle I_{\omega}\left(  \mathbf{r}\right)  \right\rangle
\approx\widehat{\mathbf{r}}\frac{Q^{2}\rho_{0}^{2}}{2}\left(  f_{coh}\left(
r;\omega\right)  +f_{td}\left(  r;\omega\right)  \right)  \text{.}%
\end{equation}
where $f_{coh}$ and $f_{td}$ are given by (\ref{2Dfcoh}) and (\ref{2Dftotdif})
respectively. The average intensity in 3D is approximated well by%
\begin{equation}
\left\langle I_{\omega}\left(  \mathbf{r}\right)  \right\rangle \approx
\frac{Q^{2}\rho_{0}^{2}}{2}\frac{1}{16\pi^{2}r\ell_{f}}\left(  \frac{\ell_{f}%
}{r}e^{-r/\ell_{f}}+3\right)  \text{.} \label{AvInt3D}%
\end{equation}
We obtain these expressions from our first-principles calculations that enable
us to study not only the ballistic and diffusive limits, but also the
crossover regime when $r/\ell_{f}\approx1$. From this we observe that we can
approximate the average intensity well by just the coherent and diffusive
contributions. Furthermore, we saw that already at $r/\ell_{f}\approx0.3$ the
diffusive intensity is higher than the coherent intensity. This does not mean
that when a pulsed source is used, we should see signs of the crossover to the
diffusive regime at this point, because the diffuse peak is much broader than
the coherent peak, so this crossover point is at larger values of $r/\ell_{f}%
$, as was previously reported \cite{zj1}. Obviously, our present model system
has been assumed to be boundless. In a finite slab geometry boundary
scattering, which is beyond the scope of this study, would of course affect
the results.

\subsection{Energy density}

To derive a first-principles expression for the diffusion constant from Fick's
Law (\ref{Fick1}), we still have to calculate the average energy density given
by%
\begin{align}
\left\langle W_{\omega}\left(  \mathbf{r}\right)  \right\rangle  &
=\frac{Q^{2}\rho_{0}}{4}\left(  \left\langle \left\vert \mathbf{\nabla
}G\left(  \mathbf{r},\mathbf{r}^{\prime}=0;\omega\right)  \right\vert
^{2}\right\rangle \right. \nonumber\\
&  \text{ \ \ \ \ \ \ }\left.  +\left\langle \omega^{2}c^{-2}\left(
\mathbf{r}\right)  \left\vert G\left(  \mathbf{r},\mathbf{r}^{\prime}%
=0;\omega\right)  \right\vert ^{2}\right\rangle \right)  \text{.}
\label{AvWinG}%
\end{align}
The first term is the average potential energy density and the second term
corresponds to the kinetic energy.

We start with the potential energy term in 2D and 3D. We define%
\begin{equation}
\left\langle \left\vert \mathbf{\nabla}G\left(  \mathbf{r},\mathbf{r}^{\prime
}=0;\omega\right)  \right\vert ^{2}\right\rangle =\overset{\prime\prime}{\Pi
}\left(  r;\omega\right)  \text{.}%
\end{equation}
The Fourier transform of $\overset{\prime\prime\text{ \ }}{\Pi_{0}}%
$($=\left\vert \left\langle \mathbf{\nabla}G\right\rangle \right\vert ^{2}$)
diverges, which means that we can not use the same procedure as we used for
the intensity and the flux. According to the Bethe-Salpeter equation%
\begin{align}
\overset{\prime\prime}{\Pi}\left(  r;\omega\right)   &  =\overset{\prime
\prime\text{ \ }}{\Pi_{0}}\left(  r;\omega\right)  +\Pi_{0}^{-1}\left(
k=0;\omega\right) \nonumber\\
&  \times%
%TCIMACRO{\dint }%
%BeginExpansion
{\displaystyle\int}
%EndExpansion
d^{d}\mathbf{r}_{1}\overset{\prime\prime\text{ \ }}{\Pi_{0}}\left(
r_{1};\omega\right)  \Pi\left(  \left\vert \mathbf{r-r}_{1}\right\vert
;\omega\right)  \text{.}%
\end{align}
This integral diverges as well because of the strong singularities in
$\overset{\prime\prime\text{ \ }}{\Pi_{0}}$ (also when the gradient is
calculated in the 2D\ case). When averaging, scatterers are effectively moved
around the medium and for every configuration the contribution to the total
average response is calculated. However, because of the stronger singularities
in $\overset{\prime\prime\text{ \ }}{\Pi_{0}}$ (remember that every scatterer
becomes a new source of spherical waves) this is not possible when the
receiver position coincides with a scatterer position. The reason for this is
the point receiver assumption and the far field scattering approximation. We
can circumvent this problem by omitting a small volume/area around
$\mathbf{r}_{1}$ with radius of approximately one wavelength. This slightly
modifies the probability distribution function form \textquotedblleft
completely random\textquotedblright\ to \textquotedblleft
non-overlapping\textquotedblright\ (with the receiver) in order to avoid the
divergencies. We then find that $\overset{\prime\prime}{\Pi}$ is given by:%
\begin{equation}
\overset{\prime\prime}{\Pi}\left(  r;\omega\right)  =\left\vert \kappa
_{e}\right\vert ^{2}\Pi\left(  r;\omega\right)  \text{.} \label{PidaccAppr}%
\end{equation}
In principle, our original expression for $\Pi$ should now be multiplied by a
factor $\exp\left(  -r_{o}/\ell_{f}\right)  $, where $r_{o}$ is the radius of
omission, so as long as the mean free path is longer then a few wavelengths
omitting this small volume does not influence the results. Furthermore, even
if scattering is strong and the mean free path is of the order of the
wavelength, this factor will not be of importance.

The second term of Eq. (\ref{AvWinG}), the kinetic energy, can be split:%
\begin{align}
&  \left\langle \omega^{2}c^{-2}\left(  \mathbf{r}\right)  \left\vert G\left(
\mathbf{r},\mathbf{r}^{\prime}=0;\omega\right)  \right\vert ^{2}\right\rangle
\nonumber\\
&  =\kappa_{0}^{2}\Pi\left(  r;\omega\right)  -\left\langle V\left(
\mathbf{r};\omega\right)  \left\vert G\left(  \mathbf{r},\mathbf{r}^{\prime
}=0;\omega\right)  \right\vert ^{2}\right\rangle \text{.}%
\end{align}
Now the condition that the scatterer position can not coincide with the
receiver position ensures that the second term vanishes, due to the step
function in the potential (\ref{ScatPot}). We can thus just disregard this term.

In 1D it is straightforward to prove that%
\begin{equation}
\overset{\prime\prime}{\Pi}\left(  \left\vert x\right\vert ;\omega\right)
=\left\vert \kappa_{e}\right\vert ^{2}\Pi\left(  \left\vert x\right\vert
;\omega\right)  \text{,}%
\end{equation}
always holds. We have to impose the condition the the receiver can not
coincide with a scatterer to make sure that%
\begin{equation}
\left\langle \omega^{2}c^{-2}\left(  x\right)  \left\vert G\left(
x,x^{\prime}=0;\omega\right)  \right\vert ^{2}\right\rangle =\kappa_{0}^{2}%
\Pi\left(  \left\vert x\right\vert ;\omega\right)  \text{.}%
\end{equation}
Only under the restrictions mentioned here, the averaged energy density in 1D,
2D and 3D can be expressed as being proportional to the intensity:%
\begin{equation}
\left\langle W_{\omega}\left(  \mathbf{r}\right)  \right\rangle =\frac
{1}{2\rho_{0}}\left(  \left\vert \kappa_{e}\right\vert ^{2}+\kappa_{0}%
^{2}\right)  \left\langle I_{\omega}\left(  \mathbf{r}\right)  \right\rangle
\text{.} \label{WpropI}%
\end{equation}
and this thus means that only the gradient of the energy density is
well-defined in the 1D and 2D cases.

\section{The diffusion constant}

Using the Bethe-Salpeter equation with the Ward identity we find expressions
for the average energy flux (\ref{AvFlux1D}-\ref{AvFlux3D}), the (gradient of)
the average intensity (\ref{GrAvInt1D}-\ref{AvInt3D}). The average energy
density is just proportional to the average intensity, Eq. (\ref{WpropI}).
When $r/\ell_{f}\gg1$ we expect (\ref{Fick1}) to hold and as the gradient of
the average energy density and the average flux are now known we find an
expression for the diffusion constant from (\ref{Fick1}). This means that the
diffusion constant can be written as%
\begin{equation}
D\left(  \omega\right)  =\frac{1}{d}c_{eff}\left(  \omega\right)  \ell
_{f}\left(  \omega\right)  \text{,}%
\end{equation}
where in the 1D and 3D case%
\begin{equation}
c_{eff}\left(  \omega\right)  =c_{0}\frac{2\kappa_{r}\left\vert \kappa
_{0}\right\vert }{\kappa_{r}^{2}+1/\left(  2\ell_{f}\right)  ^{2}+\kappa
_{0}^{2}}\text{,} \label{ceff1D3D}%
\end{equation}
and in the 2D case%
\begin{equation}
c_{eff}\left(  \omega\right)  =c_{0}\frac{2\kappa_{r}\left\vert \kappa
_{0}\right\vert }{\kappa_{r}^{2}+1/\left(  2\ell_{f}\right)  ^{2}+\kappa
_{0}^{2}}g\left(  2\kappa_{r}\ell_{f}\right)  \text{,}%
\end{equation}
where $g\left(  2\kappa_{r}\ell_{f}\right)  $ is given by (\ref{g2D}). The
effective transport velocity in 2D reduces to (\ref{ceff1D3D}) in the weak
scattering limit.

We can now investigate the frequency dependence of the diffusion constant for
a medium with monodisperse scatterers. We relate the scatterer density $n$ to
the average distance between scatterers ($\left\langle d_{s}\right\rangle $)
so that $n=\left\langle d_{s}\right\rangle ^{-1}$ in 1D, $n=4\pi
^{-1}\left\langle d_{s}\right\rangle ^{-2}$ in 2D and $n=3\left(  4\pi\right)
^{-1}\left\langle d_{s}\right\rangle ^{-3}$ in 3D. Let us focus on the
diffusion constant of the 2D medium. We write%
\begin{equation}
a\kappa_{e}\left(  a\kappa_{0}\right)  =\sqrt{\left(  a\kappa_{0}\right)
^{2}-\frac{4}{\pi}\left(  \frac{a}{\left\langle d_{s}\right\rangle }\right)
^{2}t_{0}\left(  a\kappa_{0}\right)  }\text{,}%
\end{equation}
so that the dimensionless property $a\kappa_{e}$ depends on the dimensionless
frequency $\kappa_{0}a$ ($=\omega a/c_{0}$) and two dimensionless model
parameters, i.e. the velocity contrast $\gamma(=c_{int}/c_{0})$ and the
average distance between scatterers in number of scatterer radii
($\left\langle d_{s}\right\rangle /a$). The real and imaginary parts of
$a\kappa_{e}$ are needed to obtain the diffusion constant%
\begin{equation}
a\kappa_{r}\left(  a\kappa_{0}\right)  =\left\vert \operatorname{Re}\left\{
a\kappa_{e}\left(  a\kappa_{0}\right)  \right\}  \right\vert \text{,}%
\end{equation}%
\begin{equation}
\frac{\ell_{f}\left(  a\kappa_{0}\right)  }{a}=\frac{1}{2\left\vert
\operatorname{Im}\left\{  a\kappa_{e}\left(  a\kappa_{0}\right)  \right\}
\right\vert }\text{.}%
\end{equation}

\begin{figure}
[ptb]
\begin{center}
\includegraphics[
height=2.6991in,
width=3.4in
]%
{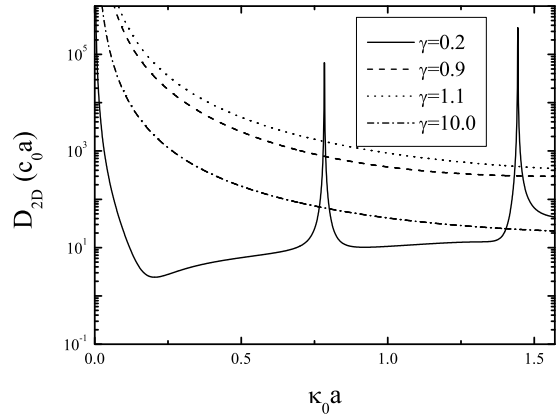}%
\caption{Diffusion constant of the 2D disordered medium in units of $c_{0}a$,
as a function of the dimensionless frequency $\kappa_{0}a$ for four different
scatterer-medium velocity ratios ($\gamma$). The scatterer density is
determined by setting $\left\langle d_{s}\right\rangle /a=10$.}%
\label{Fig2DDifCon}%
\end{center}
\end{figure}

The diffusion constant for a 2D medium is plotted in Fig. \ref{Fig2DDifCon}.
The relevant frequency range is from $\kappa_{0}a$($=\omega a/c_{0}$)$=0$ to
$\kappa_{0}a\approx\pi/2$, as for higher frequencies the isotropic scatterer
assumption is no longer valid. For the plot, the density of scatterers was
determined by setting $\left\langle d_{s}\right\rangle /a=10$, increasing this
value shifts the curves up. The shape of the curves is predominantly
determined by the mean free path. The effective transport velocity $c_{eff}$
only deviates considerably from $c_{0}$ when the scatterer velocity and the
frequency are small and the scatterer density high. For the diffusion
constants shown in the plot, this is only the case when $\gamma=0.2$. This is
also the only case that shows resonances in the relevant frequency range.
Lowering the internal velocity of the scatterers even more, would
\textquotedblleft pull in\textquotedblright\ more resonances in the relevant
frequency range. These resonances show up because of resonances in the mean
free path. When the scatterer-medium velocity ratio is increased, the mean
free path (and thus the diffusion constant) increases until the ratio is
larger than unity and it will drop again. However, increasing $\gamma$ above
$10$, will not change the diffusion constant much in the frequency range we
discuss.%
%TCIMACRO{\FRAME{ftbpFU}{3.5965in}{2.8551in}{0pt}{\Qcb{Diffusion constant of
%the 2D disordered medium in units of $c_{0}a$, as a function of the
%dimensionless frequency $\kappa_{0}a$ for four different scatterer-medium
%velocity ratios ($\gamma$). The scatterer density is determined by setting
%$\left\langle d_{s}\right\rangle /a=10$.}}{\Qlb{Fig2DDifCon}}{2ddifconb.eps}%
%width 3.5965in;  height 2.8551in;  depth 0pt;  original-width 4.1984in;
%original-height 3.325in;  cropleft "0";  croptop "1";  cropright "1";
%cropbottom "0";  filename '2DDifConb.EPS';file-properties "NPEU";}}}%
%BeginExpansion
%EndExpansion

The diffusion constants in 1D and 3D media show the same behavior. Of course,
the resonances at low velocity, are caused by the fact that all scatterers are
assumed to have equal size. When scatterer sizes (or velocities) are allowed
to vary, the resonances will be averaged out.

When the scatterer velocity is zero we obtain an impenetrable model scatterer.
This is not a useful model scatterer, as in the low frequency range the mean
free path (and thus the diffusion constant) differ considerably from the
penetrable scatterer case. The reason for this is that the limits for
$\omega\rightarrow0$ and $\gamma\rightarrow0$ do not commute, as%
\begin{equation}
\lim_{\omega\rightarrow0}\lim_{\gamma\rightarrow0}\ell_{f}=\text{constant,}%
\end{equation}
while%
\begin{equation}
\lim_{\gamma\rightarrow0}\lim_{\omega\rightarrow0}\ell_{f}=\infty\text{.}%
\end{equation}
The effect of this is that at the longer wavelengths, $\ell_{f}$ for the
impenetrable scatterer is orders of magnitude smaller than $\ell_{f}$ for
non-zero values of $\gamma$.

\section{Conclusions}

We have calculated the transport of energy and intensity in disordered 1D, 2D
and 3D (infinite) media emitted by a monochromatic source. Using the ladder
approximation to the Bethe-Salpeter equation we explicitly show that the total
intensity is well approximated by the sum of the coherent and the fully
developed diffuse wave field, for all source-receiver distances. Energy
transport and intensity propagation in 2D disordered systems shows
interestingly different behavior compared to the 3D case: In 2D, the average
energy flux depends on the mean free path and the effective transport velocity
depends differently in terms of the scattering parameters. The (gradient of
the) intensity as a function of the source-receiver distance, on the other
hand, behaves similarly in the 2D and the 3D case. The monochromatic source
enables us to investigate the frequency dependence due to finite size
scatterers on the macroscopic diffusion constant. For a monodisperse
distribution of scatterers shape resonances show up in the relevant frequency
range for low internal scatterer velocities ($\gamma$ small). In this
frequency range (where scattering is expected to be isotropic) the dependence
of the scattering properties on frequency can not be neglected. This means
that descriptions of broadband pulse propagation through these media should in
principle incorporate both frequency dependent and multiple scattering
effects. The development of a workable Ward identity in this case remains a
challenge, however. Finally, we want point out that our model describes
transport of scalar acoustic waves but results can be extended and many
conclusions should also apply to vector wave fields random media.

\section*{Acknowledgments}

We thank Mauro Ferreira and Kees Wapenaar for discussions. This work is part
of the research program of the \textquotedblleft Stichting Technische
Wetenschappen\textquotedblright\ (STW) and the \textquotedblleft Stichting
Fundamenteel Onderzoek der Materie\textquotedblright\ (FOM) which is
financially supported by the \textquotedblleft Nederlandse Organisatie voor
Wetenschappelijk Onderzoek\textquotedblright\ (NWO).

\appendix{}

\section{Energy and intensity in 2D}

In this appendix we derive the configuration-averaged intensity and energy
flux in a disordered 2D\ medium. Starting point is the 2D\ Green function
propagator%
\begin{equation}
G\left(  r;\omega\right)  =\left\{
\begin{array}
[c]{c}%
-\frac{i}{4}H_{0}^{(1)}\left(  \left(  \kappa_{r}+i/\left(  2\ell_{f}\right)
\right)  r\right)  \text{ \ \ if \ }\omega>0\\
\frac{i}{4}H_{0}^{(2)}\left(  \left(  \kappa_{r}-i/\left(  2\ell_{f}\right)
\right)  r\right)  \text{ \ \ \ \ if \ }\omega<0
\end{array}
\right.  \text{.}%
\end{equation}
We use the properties%
\begin{align}
&  H_{0}^{(2)}\left(  \left(  \kappa_{r}-i/\left(  2\ell_{f}\right)  \right)
r\right) \nonumber\\
&  =H_{0}^{(1)}\left(  \left(  -\kappa_{r}+i/\left(  2\ell_{f}\right)
\right)  r\right)  \text{,}%
\end{align}
and%
\begin{align}
&  H_{0}^{(1)}\left(  \left(  \pm\kappa_{r}+i/\left(  2\ell_{f}\right)
\right)  r\right) \nonumber\\
&  =-i\frac{2}{\pi}K_{0}\left(  \left(  \mp i\kappa_{r}+1/\left(  2\ell
_{f}\right)  \right)  r\right)  \text{,}%
\end{align}
to express the Hankel functions ($H_{0}^{(j)}$) in terms of modified Bessel
function of the second kind ($K_{0}$). The Fourier transform of the coherent
intensity%
\begin{equation}
\Pi_{0}\left(  k;\omega\right)  =2\pi%
%TCIMACRO{\dint \limits_{0}^{\infty}}%
%BeginExpansion
{\displaystyle\int\limits_{0}^{\infty}}
%EndExpansion
\left\vert G\left(  r;\omega\right)  \right\vert ^{2}J_{0}\left(  kr\right)
\text{,}%
\end{equation}
is then obtained from Ref. \cite{gr1}\ and using properties of the associated
Legendre polynomials \cite{as1}\ as%
\begin{equation}
\Pi_{0}\left(  k;\omega\right)  =\frac{\ell_{f}^{2}}{\pi}\frac{1}{1+\left(
2\kappa_{r}\ell_{f}\right)  ^{2}}\frac{P_{-1/2}^{-1/2}\left(  u\right)
}{P_{1/2}^{-1/2}\left(  u\right)  }\text{,}%
\end{equation}
with%
\begin{equation}
u=\frac{1-\left(  2\kappa_{r}\ell_{f}\right)  ^{2}}{1+\left(  2\kappa_{r}%
\ell_{f}\right)  ^{2}}+\frac{2k\ell_{f}}{1+\left(  2\kappa_{r}\ell_{f}\right)
^{2}}\text{.}%
\end{equation}
This is can be rewritten as%
\begin{equation}
\Pi_{0}\left(  k;\omega\right)  =\frac{\ell_{f}^{2}}{\pi}\frac{\arcsin\left(
\frac{\sqrt{\left(  2\kappa_{r}\ell_{f}\right)  ^{2}-\left(  k\ell_{f}\right)
^{2}}}{\sqrt{1+\left(  2\kappa_{r}\ell_{f}\right)  ^{2}}}\right)  }%
{\sqrt{1+\left(  k\ell_{f}\right)  ^{2}}\sqrt{\left(  2\kappa_{r}\ell
_{f}\right)  ^{2}-\left(  k\ell_{f}\right)  ^{2}}}\text{.}%
\end{equation}
$\Pi_{0}\left(  k;\omega\right)  $ is real, continuous and differentiable for
all (real) $k\geq0$.

The flux in the 2D system is given by%
\begin{equation}
\left\langle \mathbf{n\cdot F}_{\omega}\left(  \mathbf{r}\right)
\right\rangle =\frac{Q^{2}\rho_{0}\omega}{2}\frac{C}{r}\mathbf{n}\cdot
\widehat{\mathbf{r}}\text{,}%
\end{equation}
where $C$ is the constant to be calculated:%
\begin{equation}
C=\frac{%
%TCIMACRO{\dint \limits_{0}^{\infty}}%
%BeginExpansion
{\displaystyle\int\limits_{0}^{\infty}}
%EndExpansion
dr\operatorname{Im}\left\{  G\left(  r;\omega\right)  \partial_{r}G^{\ast
}\left(  r;\omega\right)  \right\}  r^{2}}{\Pi_{0}^{-1}\left(  k=0;\omega
\right)  \left.  \partial_{k}^{2}\Pi_{0}\left(  k;\omega\right)  \right\vert
_{k=0}}\text{.}%
\end{equation}
The term in the denominator is easily obtained%
\begin{align}
&  \Pi_{0}^{-1}\left(  k=0;\omega\right)  \left.  \partial_{k}^{2}\Pi
_{0}\left(  k;\omega\right)  \right\vert _{k=0}\nonumber\\
&  =-\ell_{f}^{2}\left(  1-\frac{1}{\left(  2\kappa_{r}\ell_{f}\right)  ^{2}%
}+\frac{1}{2\kappa_{r}\ell_{f}\arctan\left(  2\kappa_{r}\ell_{f}\right)
}\right)  \text{.}%
\end{align}
The solution to the integral
\begin{align}
&
%TCIMACRO{\dint \limits_{0}^{\infty}}%
%BeginExpansion
{\displaystyle\int\limits_{0}^{\infty}}
%EndExpansion
dr\operatorname{Im}\left\{  G\left(  r;\omega\right)  \partial_{r}G^{\ast
}\left(  r;\omega\right)  \right\}  r^{2}\nonumber\\
&  =-\frac{sgn\left(  \omega\right)  }{4\pi^{2}\ell_{f}}%
%TCIMACRO{\dint \limits_{0}^{\infty}}%
%BeginExpansion
{\displaystyle\int\limits_{0}^{\infty}}
%EndExpansion
drr^{2}\operatorname{Im}\left\{  \left(  i\kappa_{r}+1/\left(  2\ell
_{f}\right)  \right)  \right. \nonumber\\
&  \left.  \times K_{0}\left(  \left(  -i\kappa_{r}+1/\left(  2\ell
_{f}\right)  \right)  r\right)  K_{1}\left(  \left(  i\kappa_{r}+1/\left(
2\ell_{f}\right)  \right)  r\right)  \right\} \nonumber\\
&  =-\frac{sgn\left(  \omega\right)  }{4\pi^{2}\ell_{f}^{2}}\operatorname{Im}%
\left\{  2\frac{\left(  1+i2\kappa_{r}\ell_{f}\right)  ^{2}}{\left(
1-i2\kappa_{r}\ell_{f}\right)  ^{4}}\right. \nonumber\\
&  \left.  \times F\left(  2,2;3;1-\frac{\left(  1+i2\kappa_{r}\ell
_{f}\right)  ^{2}}{\left(  1-i2\kappa_{r}\ell_{f}\right)  ^{2}}\right)
\right\}  \text{,}%
\end{align}
can be found from Ref. \cite{gr1}. However, the proper solution (on the right
Riemann sheet) needs to be chosen in order to simplify the hypergeometric
series $F$. One can check numerically that%
\begin{align}
&
%TCIMACRO{\dint \limits_{0}^{\infty}}%
%BeginExpansion
{\displaystyle\int\limits_{0}^{\infty}}
%EndExpansion
dr\operatorname{Im}\left\{  G\left(  r;\omega\right)  \partial_{r}G^{\ast
}\left(  r;\omega\right)  \right\}  r^{2}\nonumber\\
&  =-\frac{sgn\left(  \omega\right)  }{4\pi^{2}\ell_{f}^{2}}\arctan\left(
2\kappa_{r}\ell_{f}\right) \nonumber\\
&  \times\left(  1-\frac{1}{\left(  2\kappa_{r}\ell_{f}\right)  ^{2}}+\frac
{1}{2\kappa_{r}\ell_{f}\arctan\left(  2\kappa_{r}\ell_{f}\right)  }\right)
\text{.}%
\end{align}
Hence, $C$ is given by
\begin{equation}
C=\frac{sgn\left(  \omega\right)  }{4\pi^{2}}\arctan\left(  2\kappa_{r}%
\ell_{f}\right)  \text{.}%
\end{equation}

The intensity is proportional to the propagator $\Pi\left(  r;\omega\right)
$, expressed in terms of $\Pi_{0}\left(  k;\omega\right)  $ by the Fourier
transform (\ref{FTInt}). Only the gradient of the intensity is a well-defined
property, so we calculate%
\begin{equation}
\nabla\Pi\left(  r;\omega\right)  =-\widehat{\mathbf{r}}%
%TCIMACRO{\dint \limits_{0}^{\infty}}%
%BeginExpansion
{\displaystyle\int\limits_{0}^{\infty}}
%EndExpansion
\frac{dk}{2\pi}\frac{k^{2}J_{1}\left(  kr\right)  }{\Pi_{0}^{-1}\left(
k;\omega\right)  -\Pi_{0}^{-1}\left(  k=0;\omega\right)  }\text{.}%
\end{equation}
This part contains both the coherent and the scattered intensity. As the
coherent intensity is known, we focus on the scattered intensity by
calculating%
\begin{equation}
\nabla\Pi_{sc}\left(  r;\omega\right)  =-\widehat{\mathbf{r}}%
%TCIMACRO{\dint \limits_{0}^{\infty}}%
%BeginExpansion
{\displaystyle\int\limits_{0}^{\infty}}
%EndExpansion
\frac{dk}{2\pi}J_{1}\left(  kr\right)  k^{2}\Pi_{sc}\left(  k;\omega\right)
\text{,}\label{GradIntDiff2D}%
\end{equation}
with%
\begin{equation}
\Pi_{sc}\left(  k;\omega\right)  =\frac{\Pi_{0}^{-1}\left(  k=0;\omega\right)
\Pi_{0}\left(  k;\omega\right)  }{\Pi_{0}^{-1}\left(  k;\omega\right)
-\Pi_{0}^{-1}\left(  k=0;\omega\right)  }\text{.}%
\end{equation}
(\ref{GradIntDiff2D}) is the integral to calculate numerically when we want to
calculate the gradient of the multiply scattered intensity. $\Pi_{sc}\left(
k;\omega\right)  $ is a monotonically decaying function with a maximum at
$k=0$, that vanishes as $k\rightarrow\infty$. As the Bessel function is also
decaying with $r$, we know that for $r/\ell_{f}\gg1$%
\begin{equation}
\nabla\Pi_{td}\left(  r;\omega\right)  =-\widehat{\mathbf{r}}\frac{\Pi
_{sc}\left(  k=0;\omega\right)  }{2\pi r}\text{.}%
\end{equation}
and%
\begin{equation}
\nabla\Pi_{td}\left(  r\right)  =-\widehat{\mathbf{r}}\frac{\arctan\left(
2\kappa_{r}\ell_{f}\right)  }{\pi^{2}2\kappa_{r}\ell_{f}}\frac{1}{r}%
g^{-1}\left(  2\kappa_{r}\ell_{f}\right)  \text{,}%
\end{equation}
where%
\begin{equation}
g\left(  2\kappa_{r}\ell_{f}\right)  =1-\frac{1}{\left(  2\kappa_{r}\ell
_{f}\right)  ^{2}}+\frac{1}{2\kappa_{r}\ell_{f}\arctan\left(  2\kappa_{r}%
\ell_{f}\right)  }\text{.}%
\end{equation}
$td$ stands for \textquotedblleft totally diffusive\textquotedblright.

\end{document}